\documentclass{article}
\pdfoutput=1

\usepackage{arxiv}

\usepackage[utf8]{inputenc} 
\usepackage[T1]{fontenc}    
\usepackage{hyperref}       
\usepackage{url}            
\usepackage{booktabs}       
\usepackage{amsfonts}       
\usepackage{nicefrac}       
\usepackage{microtype}      
\usepackage{lipsum}
\usepackage{caption}
\usepackage{graphicx}
\usepackage{amsmath}
\setcitestyle{numbers}

\title{GHOST SPECTROSCOPY WITH CLASSICAL CORRELATED AMPLIFIED SPONTANEOUS EMISSION PHOTONS EMITTED BY AN ERBIUM-DOPED FIBER AMPLIFIER}

\author{
  Patrick Jannassek, 
 Andreas Herdt, 
  S\'{e}bastien Blumenstein, and
 Wolfgang Els\"a{\ss}er $^{*}$\\
 \\
  Institute of Applied Physics\\
  Technische Universit\"{a}t Darmstadt\\
  Schlossgartenstrasse 7, 64289 Darmstadt, Germany \\
 \newline 
 \\ \texttt{$^{*}$corresponding author: elsaesser@physik.tu-darmstadt.de}\\
}

\begin{document}
\maketitle

\begin{abstract}
We demonstrate wavelength-wavelength correlations of classical broad-band amplified spontaneous emission (ASE) photons emitted by an erbium-doped fiber amplifier (EDFA) in a wavelength regime around 1530\,nm. We then apply these classical correlated photons in the framework of a real-world ghost spectroscopy experiment at a wavelength of 1533\,nm to acetylene (${\text{C}_{2}\text{H}_{2}}$) reproducing the characteristic absorption features of the C-H stretch and rotational bands. This proof-of-principle experiment confirms the generalization of an ASE source concept offering an attractive light source for classical ghost spectroscopy. It is expected that this will enable further disseminating ghost modality schemes by exploiting classical correlated photons towards applications in chemistry, physics and engineering. 
\end{abstract}


\section{Introduction}
Ghost Modalities (GM) as Ghost Imaging \cite{Sergienko:17, Boyd:112, Padgett:16, Shapiro:10, Padgett:17, Pittmann:95, Bennink2002,Zhang:05}, Temporal Ghost Imaging \cite{Genty16,Devaux:16,Devaux:17,Lantz2017}, Ghost Spectroscopy \cite{Valencia2003,Janassek:18,Genty:18} and Ghost Polarimetry \cite{JanassekOL:18} are photon correlation metrology techniques and they represent actually an interesting topic in quantum optics. Ghost Imaging (GI) has been the first GM, realized for the first time in 1995 with entangled photons generated by Spontaneous Parametric Down Conversion (SPDC) \cite{Pittmann:95,Valencia2003}, thus twin photons in the quantum mechanical sense \cite{Pittmann:95}. This quantum imaging modality exploits photon correlations for the image construction or reconstruction, where one photon of an (entangled) pair interacts with the object to be imaged and the experimentally determined correlation with the second photon yields the image, therefore it is called correlated two-photon imaging. Thus, a ghost image is obtained by correlating the light beam reflected or transmitted by an object with a highly correlated reference beam, which itself, however, had never interacted with the object \cite{Shih2005,Shih2007}. After the experimental demonstration of classical GI in 2004 \cite{Bennink2002,Ferri2005,JMOGatti2006} it became clear, that GI cannot be understood as an exclusive quantum effect based on the two photon interference capability of entangled light. Originally, classical GI was based on so-called pseudo-thermal light, generated by a spatially randomly modulated coherent laser beam (the so-called Arecchi or Martienssen lamp: a combination of a rotating diffuser and laser light) \cite{MartienssenAJP:64, Arecchi:65, Arecchi:66}. This light beam is subsequently divided by a beam splitter into two beams which are mutually highly correlated and exhibit high spatial intensity-intensity correlations with two-photon interference capability, a fact well-known since the fundamental Hanbury-Brown \& Twiss experiment \cite{HBT:56}. The capability of classical thermal light to mimic entanglement properties led to numerous experimental and theoretical investigations \cite{Meyers2011, Aspden:15,LugiatoItalia:13, Shapiro:10,Genovese:16}. Later on, GI has been also realized with a so-called true thermal light source, a filtered neon discharge lamp \cite{Chen_OL:09}, or even with the sun \cite{Liu_OL:14} as light source. In 2017, an ultra-miniaturized semiconductor-based broad-area superluminescent diode (BA-SLD) has been proposed and introduced to the field as a new compact light
source for classical ghost imaging having intrinsically all requirements of a GI light source \cite{Hartmann:17}.

Very recently, the classical GI concept based on the exploitation of spatial correlation was transferred into the spectral wavelength domain \cite{Janassek:18}, thus realizing for the first time a ghost spectroscopy (GS) experiment with classical thermal photons by exploiting spectral wavelength-wavelength correlations of spectrally broad-band amplified spontaneous emission (ASE) light emitted by an ultra-compact, miniaturized, optoelectronic, semiconductor-based Quantum Dot Superluminescent Diode (SLD) \cite{Janassek:18}. In the spirit of the Hanbury-Brown \& Twiss experiment \cite{HBT:56}, this light showed up to exhibit a spectral second order correlation coefficient of two, thus spectral photon bunching, one of the key requirements of ghost spectroscopy. For the demonstration of the applicability of this superluminescent diode light source for ghost spectroscopy with classical light, a “real world“ absorption spectroscopy experiment had been conceived at liquid chloroform, resulting in the central second order correlations coefficient $g^{(2)}(\tau=0, \lambda_{ref})$ as a function of the reference wavelength $\lambda_{ref}$, thus the ghost spectrum, which exhibited all the characteristic absorption fingerprints $S(\lambda)$ of chloroform at $1214$\,nm \cite{Janassek:18}. This so obtained ghost absorption spectrum has been consequently interpreted as the analogue to the ghost image, however, there in the spatial domain. This first demonstration of ghost spectroscopy with classical thermal light in analogy to GI closed a gap in the experimental photon correlation modalities. The realization of GS with thermal light has been enabled by solving the challenges of having an extremely high time resolution $\tau_{\text{measure}}$ ($\tau_{\text{measure}} \ll \tau_{\text{coherence}}$) for the measurements of intensity correlations of spectrally broad-band light emitted by SLDs with a coherence time $\tau_{\text{coherence}} = 1/\Delta \nu $. Only by exploiting interferometric Two-Photon Absorption (TPA) detection \cite{Boitier:09, Hartmann:15}
it has been possible to access the ultra-short correlation time scales in the 10 femtosecond range. The second challenge has been the difficulty to find a light source, emitting broad-band light and exhibiting the requested wavelength correlations to enable ghost spectroscopy.

Here, we demonstrate that this concept of the exploitation of classical correlated photons in the spectral domain can be generalized to amplified spontaneous emission sources in general \cite{Desurvire:89, Desurvire:94}. We prove this by performing a GS experiment at acetylene (${\text{C}_{2}\text{H}_{2}}$) with amplified spontaneous emission light from a continuous-wave erbium-doped fiber amplifier (EDFA) at $1530$\,nm. We would like emphasizing that it is not our intention to compete with high resolution spectroscopy \cite{Arita:08, Wagner:09} but rather demonstrating Ghost Spectroscopy with spectrally correlated classical photons from a novel exploited ASE source, thus contributing to the application and dissemination of correlated photon spectroscopy in the framework of ghost modalities or quantum technologies \cite{Kalashnikov2016}.

\section{Experimental set-up}

\begin{figure}[t]

\includegraphics[width=0.9\linewidth]{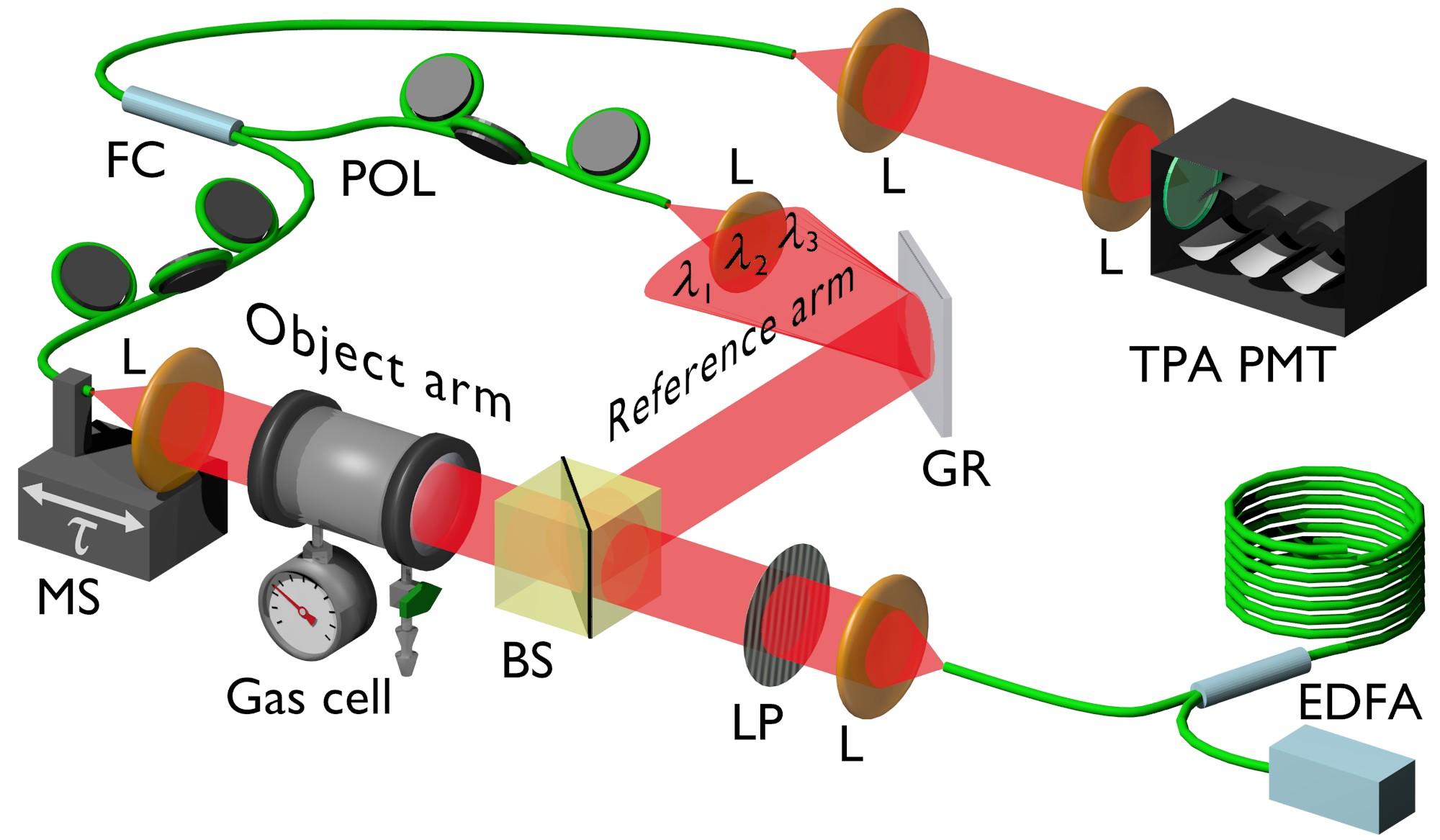}
\caption{Schematic diagram of the experimental set-up for the ghost spectroscopy experiment: The setup comprises the amplified spontaneous emission continuous wave light source of an erbium-doped fiber amplifier (EDFA), collimating and focusing achromatic lenses (L), a fixed linear polarizer (LP), a non-polarizing beamsplitter (BS), a fixed reflective diffraction grating (GR) followed by a movable fiber coupling unit in the reference arm, thus enabling wavelength resolution, a high pressure gas cell filled with ${\text{C}_{2}\text{H}_{2}}$ within the object arm, again followed by a fiber coupling unit, a single-mode fiber-based combiner (FC), a motorized linear translation stage (MS), polarization controlling elements (POL), lenses for focusing the light onto the photocathode of the PMT (Hamamatsu H7421-50) operated in TPA mode (TPA PMT) and a longpass filter (LPF, Schott RG1000) preventing visible light from entering the detector.}
\label{fig:Fig1}
\end{figure}

Figure \ref{fig:Fig1} shows a schematic drawing of the experimental setup with the EDFA on the bottom right side as the ghost spectroscopy continuous wave (CW) light source. After collimation of its fiber output, the unpolarized light beam is subsequently linear polarized by a Glan–Thompson prism resulting in a linear horizontal polarization state. Through a beam splitter, a reference beam and an object beam are generated, respectively. The object beam light passes a gas sample cell and the totally transmitted light is coupled into a single-mode fiber (SMF) serving as the spectral bucket detection branch, i.e. without any spectral resolution. In the reference beam spectral resolution can either be achieved by various, in wavelength variable tunable interference bandpass filters with a full width at half maximum (FWHM) of approximately $10$\,nm or by an Échelette grating if even higher resolution of up to $0.5$\,nm is requested. Also the reference beam is then fiber-coupled and both beams are superimposed by a fiber combiner forming an overall Mach-Zehnder-like interferometer configuration. Finally, the light is focused onto the semiconductor photocathode of a photomultiplier (PMT). The photomultiplier in use incorporates a GaAsP photocathode ($E_{\text{g}} \approx 2.04$\,eV) which has been selected regarding the source wavelengths in order to guarantee pure TPA detection. The here employed two-photon-absorption (TPA) interferometry technique for the determination of $g^{(2)}(\tau)$ goes back to the work of Boitier et al. \cite{Boitier:09} who demonstrated for the first time the photon bunching effect for black body sources with multiple THz wide optical spectra. The nonlinear TPA process requires two photons to be absorbed within a time frame given by the Heisenberg uncertainty. Hence ultra-fast intensity-intensity correlation detection $\langle I(t)I(t) \rangle$ is enabled. 
By implementing a time delay $\tau$ in the object arm via a variable optical delay line intensity auto-correlation measurements $\langle I_{\text{ref}}(t,\lambda_{\text{ref}}) I_{\text{obj}}(t+\tau)\rangle$ are performed. By reading out the photon counts from the detector output while varying the optical path of one interferometer arm using a high precision motorized linear translation stage, a TPA interferogram $I_{TPA}(\tau)$ is recorded. According to theory \cite{Boitier2013}, $I_{TPA}(\tau)$ comprises four terms: a constant one, the non-normalized second-order auto-correlation function $G^{(2)}(\tau) = \langle I(t)I(t + \tau )\rangle$ and two fast oscillating terms $F_{1}(\tau)$ and $F_{2}(\tau)$ following the center angular frequency $\omega_{0} = 2 \pi \nu_{0}$ and the frequency duplication $2\omega_{0}$ of the emitted light, respectively. Low-pass filtering of $I_{\text{TPA}}$ subsequently to an FFT allows determining $G^{(2)}(\tau)$ according to

\begin{equation}
\frac{I_{TPA}^{low-pass}(\tau)}{I_{\text{ref}}+I_{\text{obj}}}=1+2\cdot G^{(2)}(\tau)
\end{equation}

\noindent Finally, the key value for ghost modalities $g^{(2)}(\tau)$, being GI or GS, respectively,  is obtained according to \mbox{$g^{(2)}(\tau)=G^{(2)}{(\tau)} / G^{(2)}(\tau \gg\tau_{c})$} where $\tau_c$ is the first order coherence time.\\

\begin{figure} [h]
\centering
\mbox{\includegraphics[width=.3\linewidth]{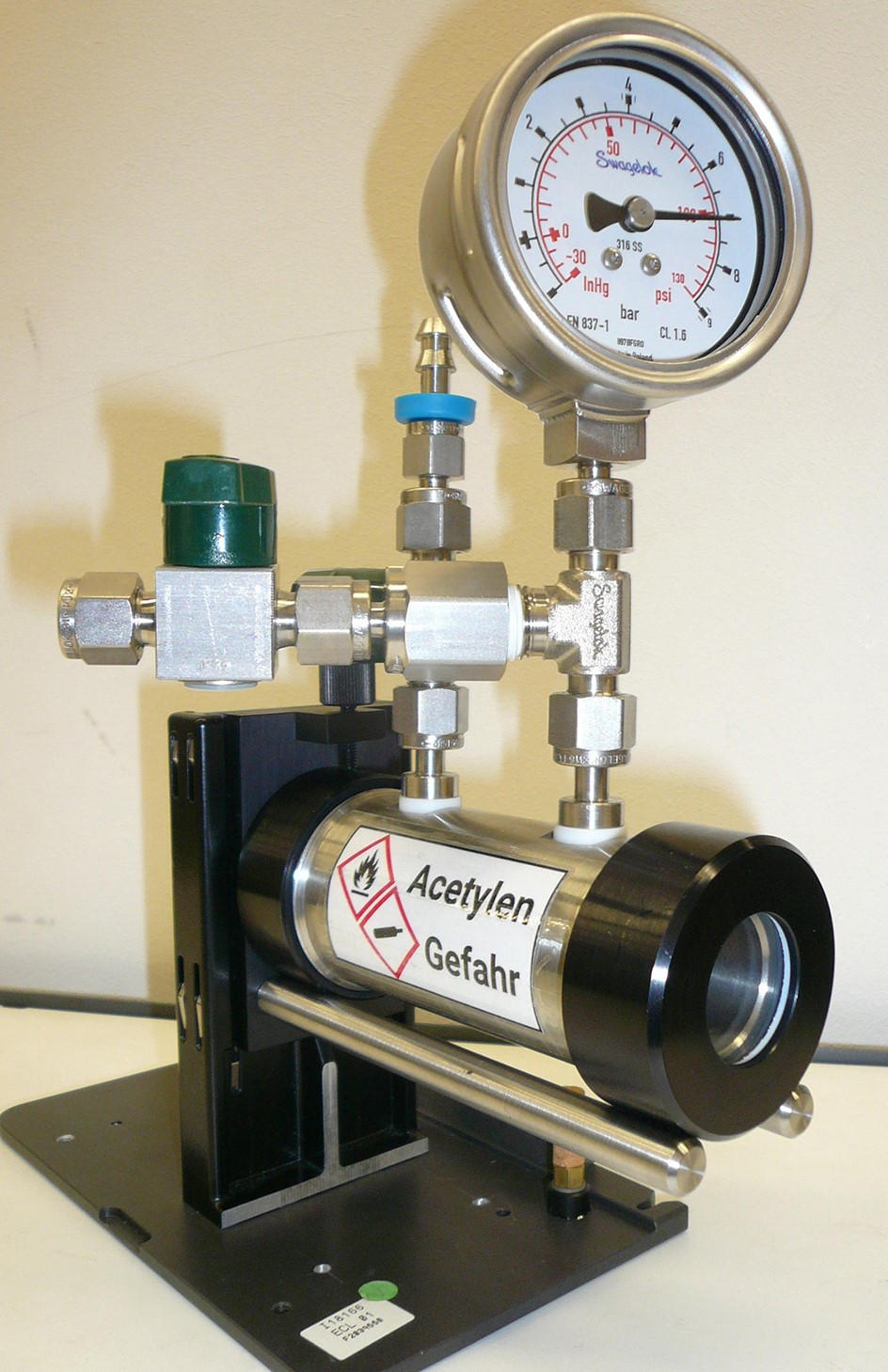}\includegraphics[width=.7\linewidth]{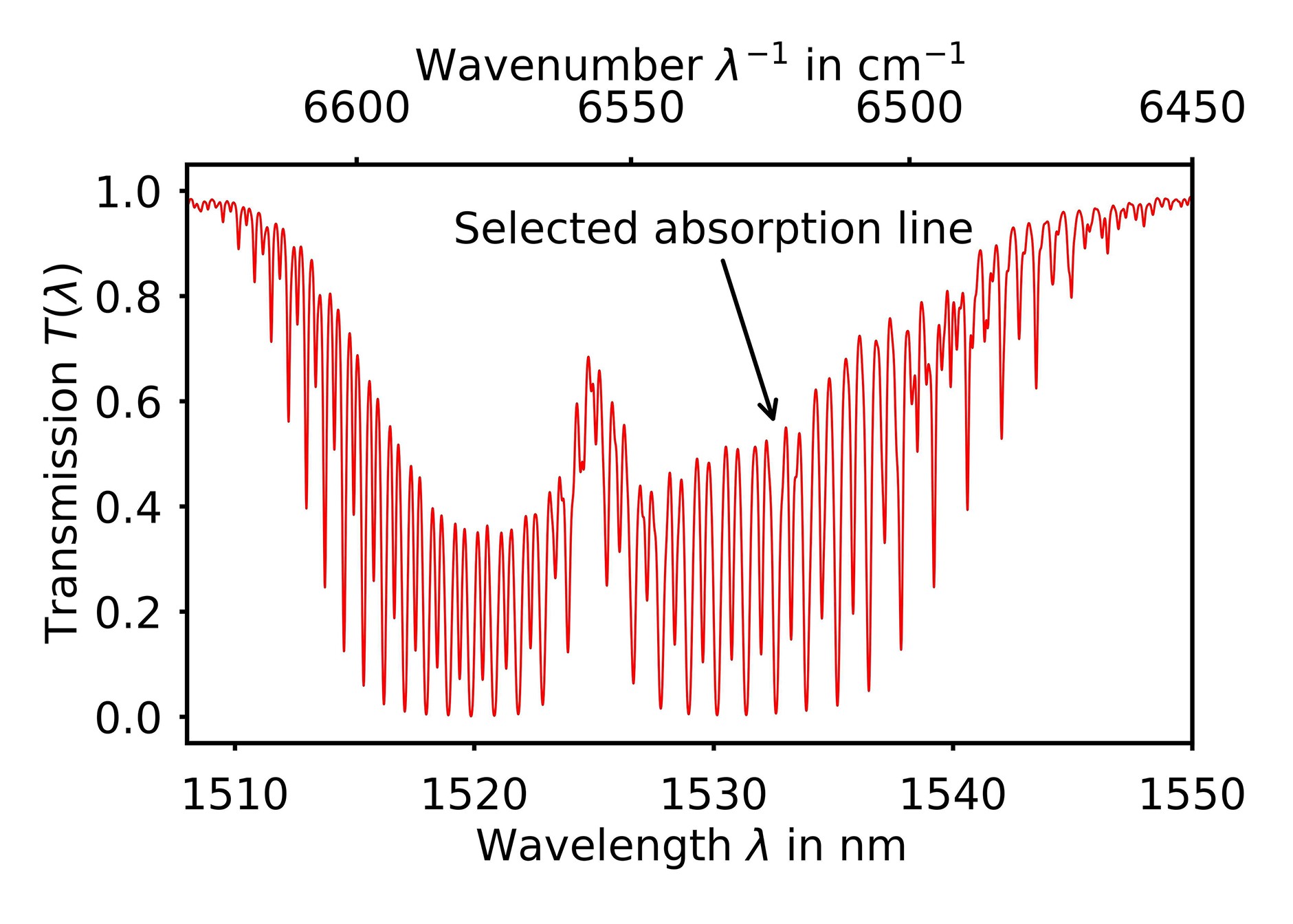}}
\caption{Left: High pressure absorption cell for the measurements and right: Conventionally measured standard absorption spectrum of acetylene around $1530$\,nm with an absorption length of $10$\,cm and an absolute pressure of $200$\,kPa.}
\label{fig:Fig2}
\end{figure}

In order to proof the concept of correlated photon spectroscopy by exploiting amplified spontaneous emission from the EDFA source \cite{Desurvire:89, Desurvire:94}, we selected acetylene (${\text{C}_{2}\text{H}_{2}}$), also called ethyne as sample gas due to the match between the emission spectrum of the EDFA and the absorption features of acetylene having an absorption band from approx. $1510$ to $1540$\,nm based on a simultaneous excitation of symmetric and asymmetric stretch bands within the CH-groups of the molecule \cite{Barrow:62} complemented by differently strong rotational excitations leading to a total of approx. 50 strong absorption lines. Under standard pressure conditions they exhibit a full width half maximum (FWHM) smaller than $0.5$\,nm. By increasing the gas pressure, broadening can be achieved either by self broadening or by cross broadening, e.g. by nitrogen. For the experiment we used a gas cell also suitable for high pressure experiments as shown in Fig. \ref{fig:Fig2} (left).  

Its volume amounts to $50$\,ml with an optical aperture of $25$\,mm and an absorption length of $100$\,mm. The optical windows consist of calcium fluoride ($\text{CaF}_{2}$) suitable for spectroscopic investigations from UV to MIR with a pressure adopted thickness of $7$\,mm. Standard spectroscopy investigations up to a pressure of $800$\,kPa showed that with increasing pressure the cross broadening leads to broader lines, however, with also decreasing absorption strength. As a compromise between absorption strength and linewidth, we have finally chosen pure acetylene with an absolute pressure of $200$\ kPa. This selection has been also supported by HITRAN modeling \cite{Gordon2017} (see later on under ghost spectroscopy result discussion).  A typical absorption spectrum measured by a Fourier Transform Interferometer (Bruker Vertex 80v) for these conditions is shown in Fig.\,\ref{fig:Fig2} (right) depicting the complex stretch and rotational vibration spectrum with more than 50 lines including the line at $1532.9$\,nm finally selected to be in the center of the investigations for the ghost spectroscopy.

\section{Experimental results and discussions: Frequency correlations of light emitted by the EDFA and ghost spectrum of acetylene}

\begin{figure}[htbp]
\centering
\mbox{\includegraphics[width=\linewidth]{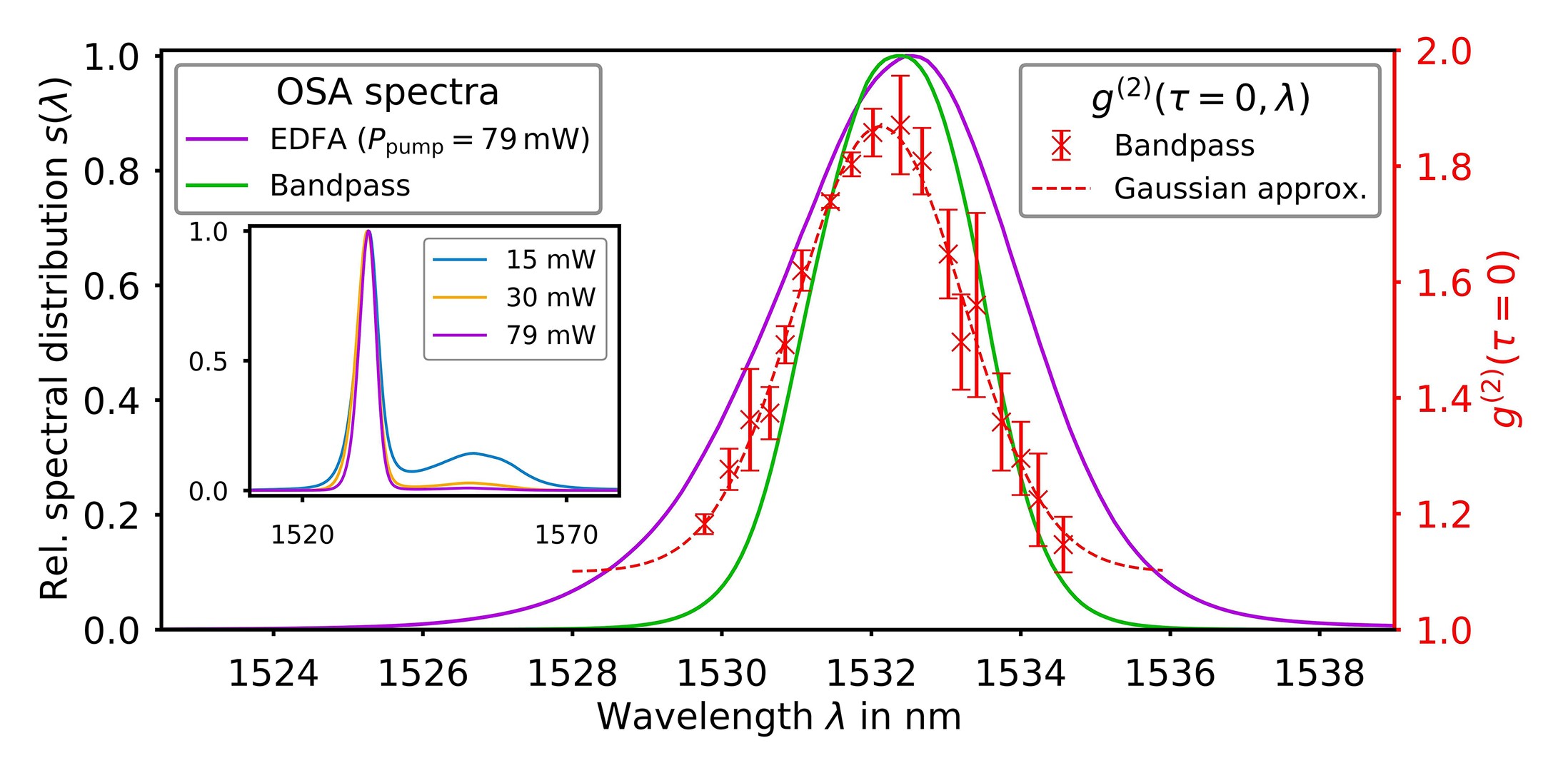}}
\caption{Measured wavelength-wavelength correlations of light emitted by the EDFA: The full optical spectrum of the EDFA (magenta solid line) has been measured by a standard optical spectrum analyzer. The selected wavelength band is also depicted (green solid line) for direct comparison with the second-order correlation data (red crosses with a dashed line based on a Gaussian fit). The error bars correspond to the standard deviation for a set of 5 measurements. The inset shows the unfiltered EDFA spectrum for three different optical pump conditions of the EDFA.}
\label{fig:Fig3}
\end{figure}

In first place, the basic requirement for a ghost spectroscopy modality is quantified, namely the wavelength-wavelength correlations of the light source. Therefore, we exploit the broad-band ASE operation of the EDFA at an optical output power of 20\,mW (corresponding to a pump power of 79\,mW) resulting in a broad-band spectrum with a central wavelength of $1533$\,nm, a FWHM of $3.8$\,nm and an obvious smaller tail up to $1570$\,nm (see Fig.  \ref{fig:Fig3}, magenta solid line).

For the investigations of the wavelength-wavelength correlations a spectral band in one interferometer arm (here the object arm) is selected by utilizing a fiber-based fixed bandpass filter with a spectral transmission as depicted by the green line in Fig.\,\ref{fig:Fig3}. Its central wavelength has been set to the maximum of the ASE spectrum of 1532.3\,nm and the spectral transmission is Gaussian-like with a FWHM of 2.5\,nm. With an additional variable fiber-based band pass filter with a FWHM of 1.4\,nm in the reference arm, the reference signal has been realized and wavelength-wavelength intensity correlations have been determined. These wavelength-wavelength intensity correlations  $g^{(2)}(\tau=0, \lambda)$ are depicted as red data points in Fig.\,\ref{fig:Fig3}. A dashed red line shows a Gaussian fit to the data.

When the wavelengths of reference and object beam coincide, a maximum $g^{(2)}(\tau=0)$ value of $1.87$ is found. We find a rapid decrease of the observed $g^{(2)}(\tau)$ values with increasing wavelength separation with a FWHM of $2.6$\,nm assuming a Gaussian shape. The $g^{(2)}(\tau=0)$ values are going down to $1.10$ in the wings at wavelengths far off by more than $4$\,nm with respect to the center. The characteristic FWHM of the $g^{(2)}(\tau=0, \lambda)$ values is nearly equal to the convolution of the FWHM of the used interference filters. Therefore, we are convinced that the observed characteristic wavelength-wavelength correlation decay values and thus the ghost spectroscopy resolution are not intrinsic to the exploited ASE but are actually due to the FWHM of the chosen interference filters used for the selection of the wavelengths (see also next paragraph).

For the ghost spectroscopy experiments, the light in the object arm has been additionally spectrally narrowed by a fiber-based spectral filter with a FWHM of $2.5$\,nm in order to restrict the rather broad ASE spectrum as already shown in the inset of Fig. \ref{fig:Fig3} for three different optical pump power conditions. The realization of a requested higher spectral resolution in the reference arm (see discussion above) has been achieved by an Échelette grating with $600\,\text{grooves}/\text{mm}$ and a blaze wavelength of $1.6\,\mu m$ (Thorlabs GR25-0616). The beam incident onto the grating has been enlarged by two cylinder lenses with focal length of  $f_{1} = 5$\,cm and $f_{2} = 9$\,cm in order to achieve a higher spectral resolution by increasing the number of illuminated grating grooves yielding finally a variable Gaussian-like wavelength reference spectrum with medium $FWHM = 0.49 \pm 0.02$\,nm. In accordance with the analogy between GI and GS, we define and exploit a ghost spectroscopy (GS) detection protocol for our experimental GS setup by starting with the GI terminology and replacing the spatial coordinate $x$ by the spectral variable $\lambda$, such that the intensity cross-correlation coefficients are now a function of the reference-arm wavelength. By varying this reference-arm wavelength, the wavelength-dependent intensity correlations, thus the ghost spectrum has been determined according to Eq. (\ref{equation1}) in the framework of a detection protocol based on GI terminology. 

\begin{equation}
g^{(2)}(\tau=0,\lambda_{\text{ref}})=\frac{\langle I_{\text{ref}}(t,\lambda_{\text{ref}}) I_{\text{obj}}(t+\tau)\rangle _t}{\langle I_{\text{ref}}(t,\lambda_{\text{ref}}) \rangle _t \langle I_{\text{obj}}(t+\tau)\rangle _t}
\label{equation1}
\end{equation}

We note that this detection protocol is based on the classical definition of the second-order Glauber correlation function which represents the most basic GI signal detection modality, now adapted in full analogy of GI towards GS.

\begin{figure*}[htbp]
\centering
\mbox{\includegraphics[width=\textwidth,height=5cm]{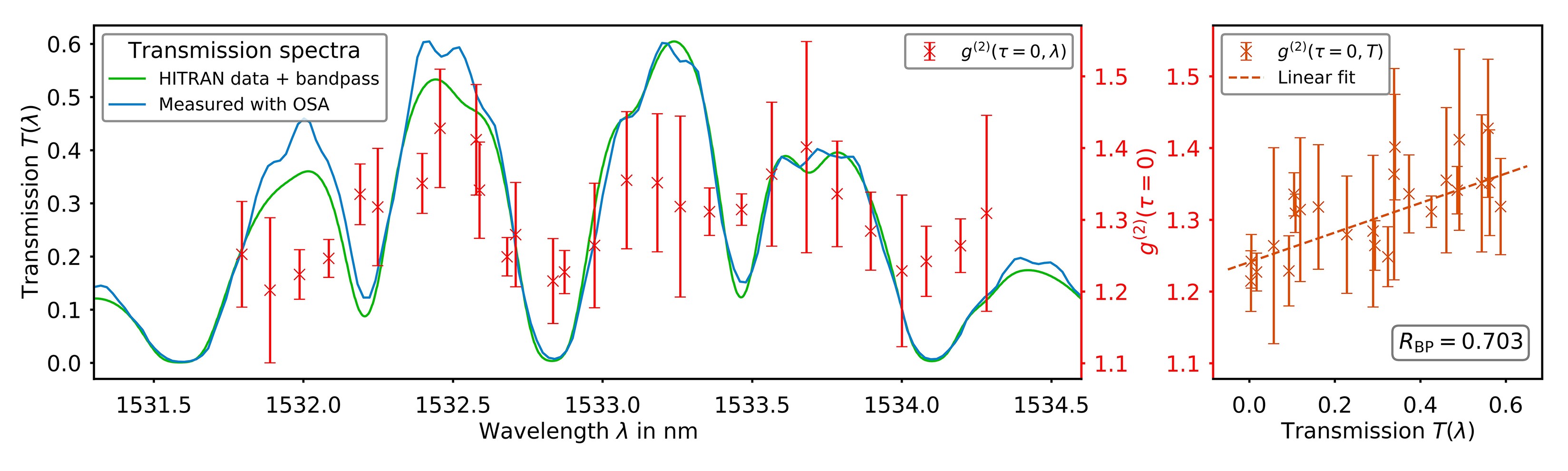}}
\caption{Left: Ghost spectrum of an absorption feature of acetylene (${\text{C}_{2}\text{H}_{2}}$) at a pressure of $200$\,kPa around $1533$\,nm measured with the set-up of Fig. \ref{fig:Fig1}. Conventionally measured standard absorption spectrum of acetylene (blue, left scale), corrected HITRAN transmission (green, left scale) and ghost spectrum (red, right scale) obtained via the spectral intensity correlations. The error bars correspond to the standard deviation of a set of 5 measurements. Right: Correlation between $g^{(2)}\left(\tau=0,\lambda_i\right)$ and $T\left(\lambda_i\right)$ evaluated in terms of the BP correlation coefficient according to Eq. (\ref{equation2}).}
\label{fig:Fig4}
\end{figure*}

Figure \ref{fig:Fig4} (left) shows the ghost spectrum in terms of \mbox{$g^{(2)}(\tau=0,\lambda_{\text{ref}})$} (red data points) together with a standard absorption spectrum of acetylene (in blue). The standard spectrum as measured by a grating spectrometer (Optical spectrum analyzer, OSA) shows clearly the central absorption feature of acetylene at $1532.9$\,nm. Towards larger and smaller wavelengths additional absorption lines are visible, however, with decreasing absorption strength according to the Gaussian envelope behavior due to the additional bandpass filter in the object arm as discussed above. The central $g^{(2)}(\tau =0,\lambda)$ values exhibit a maximum of $1.43$ at $1532.46$\,nm and a local maximum of $1.36$ at $1533.08$\,nm. In between these two maximum values, i.e. at the position of the central absorption line the normalized intensity correlations decrease to a minimum value of $g^{(2)}(\tau =0,\lambda)= 1.21$. The overall reduced second order correlation values from the maximum achievable value of two here in the spectroscopy experiment in comparison with the pure wavelength-wavelength correlation measurements \mbox{(c.f. Fig. \ref{fig:Fig3})}, which there have been nearly fully observed are due to the fact that the object arm contains a wide range of uncorrelated wavelengths resulting in an overall spectral contrast reduction. This is in analogy to the spatial contrast reduction in ghost imaging when uncorrelated spatial multi-speckle contributions are superimposed in the bucket detection thus reducing the correlation signal contrast \cite{Kuhn:16, Hartmann:16,JMOGatti2006}.

In order to evaluate the quality of the ghost spectrum and its coincidence with the real-world absorption spectroscopy results more quantitatively, the obtained GS values of $g^{(2)}\left(\tau=0,\lambda_i\right)$ are underlined with data from the HITRAN database \cite{Gordon2017} shown in green in Fig. \ref{fig:Fig4} left. The HITRAN transmission spectrum of pure acetylene has been modeled at a pressure of $200$\,kPa, a temperature of $296$\,K and for an absorption length of $10$\,cm. A corrected HITRAN spectrum $T(\lambda)$ is then obtained by multiplying the normalized emission spectrum of the EDFA with the spectral bandpass.
In order to quantify the quality of the linear correlation between the values $g^{(2)}\left(\tau=0,\lambda_i\right)$ and the corrected HITRAN transmission strength $T\left(\lambda_i\right)$, we make use of the Bravais-Pearson correlation coefficient $R_\text{BP}$ as defined according to \cite{hauke2011}

\begin{align}
R_\text{BP} &=\frac{\sum_{i=1}^n \left(g^{(2)}\left(\tau=0,\lambda_i\right)-g^{(2)}_\text{mean}\right)\cdot\left(T\left(\lambda_i\right)-\bar{T}\right)}{\sum_{i=1}^n \left(g^{(2)}\left(\tau=0,\lambda_i\right)-g^{(2)}_\text{mean}\right)^2\sum_{i=1}^n\left(T\left(\lambda_i\right)-\bar{T}\right)^2}.\label{equation2}
\end{align}

It gives a reliable measure on the confidentiality of a fit with $-1 < R_\text{BP} < 1$ and $R_\text{BP}=1$ representing full correlation. The variable $n$ represents the total number of $g^{(2)}\left(\tau=0,\lambda_i\right)$ values, while $g^{(2)}_\text{mean}$ and $\bar{T}$ represent the arithmetic means of all $g^{(2)}\left(\tau=0,\lambda_i\right)$ and $T\left(\lambda_i\right)$ respectively. The values $T\left(\lambda_i\right)$ are the HITRAN transmission strengths for each wavelength of the corresponding $g^{\left(2\right)}\left(0,\lambda_i\right)$ value. The evaluation of the Bravais-Pearson correlation coefficient in our case yields $R_\text{BP} = 0.703$, demonstrating a good linear relationship between transmission $T$ and $g^{(2)}$, which is also visualized with the linear fit curve in the right part of Fig. \ref{fig:Fig4}.

 It can be seen that the central absorption line is nicely reproduced by the intensity correlations of the ghost spectrum. On the other hand, the weaker absorption lines and spectrally narrower side lines away from the central $1532.9$\, nm line are less well reproduced. Their full width half maximum is in the order of $0.15$\,nm which can not be resolved by the $0.5$\,nm broad reference beams. Furthermore, towards the periphery location of the spectral distribution ($< 1532.5$ nm and $> 1533.5$ nm) the intensity correlations show an overall decreasing tendency due to the bandpass filter (compare with Fig. \ref{fig:Fig3} and Ref. \cite {Janassek:18}). This deterioration of the intensity correlations is due to optical power limitations reflected in an increase in the error bars of the $g^{(2)}$ values exceeding 0.1, thus restricting the measurable wavelength region. A high performance telecommunication-mature state-of-the-art EDFA with sufficiently high output power (in our case we only used a rather conservative power of up to $20$\,mW) would clearly much better perform in the ghost spectroscopy investigations.

Still, the main features of the ghost spectrum clearly show the successful spectroscopic identification of a narrow absorption line of acetylene, thus demonstrating the spectrally correlated photon concept from amplified spontaneous emission from an EDFA for ghost spectroscopy with classical photons.

\section{Summary}
We generalized the applicability of the amplified spontaneous emission concept of classical correlated photons from an erbium-doped fiber amplifier to ghost spectroscopy. After the proof of the pre-requisite of wavelength-wavelength correlations in the emitted continuous wave broad-band ASE we succeeded in performing ghost spectroscopy with a gas sample reproducing the characteristic absorption feature of acetylene at $1533$\,nm. This is the first time that a fiber-based ASE source has been applied to ghost spectroscopy. We expect that the highlights of the realized innovative scheme in the spirit of the ingredients of ghost spectroscopy with classical light with particular emphasis on the conceived source and detection schemes will further stimulate new applications of ghost modalities \cite{Genty:18} thus further fertilizing the field and allowing to develop an even deeper understanding of the experimental scheme and ghost protocols. This will open avenues for perspectives and dissemination of the ghost modality idea in entering further real-world applications \cite{LugiatoItalia:13,Genovese:16} in Chemistry, Physics and Engineering.

\section{Funding.} Deutsche Forschungsgemeinschaft (DFG) (EL105/21); Deutscher Akademischer Austauschdienst (DAAD) (Breakthroughs in Ghost Imaging).

\section{Acknowledgment.} We thank Prof. A. T. Friberg and Prof. G. Genty for discussions about GI modalities.

\bibliographystyle{unsrt}
\bibliography{manuscript}

\begin{thebibliography}{10}

\bibitem{Sergienko:17}
D.~S. Simon, G.~Jaeger, and A.~V. Sergienko.
\newblock {\em Quantum Metrology, Imaging, and Communication}.
\newblock Springer, 2017.

\bibitem{Boyd:112}
J.~H. Shapiro and R.~W. Boyd.
\newblock The physics of ghost imaging.
\newblock {\em Quantum Inf. Process.}, 11:949–993, 2012.

\bibitem{Padgett:16}
Miles Padgett, Reuben Aspden, Graham Gibson, Matthew Edgar, and Gabe Spalding.
\newblock Ghost imaging.
\newblock {\em Opt. Photon. News}, 27:38--45, 2016.

\bibitem{Shapiro:10}
Baris~I. Erkmen and Jeffrey~H. Shapiro.
\newblock Ghost imaging: from quantum to classical to computational.
\newblock {\em Adv. Opt. Photon.}, 2:405--450, 2010.

\bibitem{Padgett:17}
Miles~J Padgett and Robert~W Boyd.
\newblock An introduction to ghost imaging: quantum and classical.
\newblock {\em Phil. Trans. R. Soc. A}, 375:20160233, 2017.

\bibitem{Pittmann:95}
T.~B. Pittman, Y.~H. Shih, D.~V. Strekalov, and A.~V. Sergienko.
\newblock Optical imaging by means of two-photon quantum entanglement.
\newblock {\em Phys. Rev. A}, 52:R3429--R3432, Nov 1995.

\bibitem{Bennink2002}
R.~S. Bennink, S.~J. Bentley, and R.~W. Boyd.
\newblock "\uppercase{T}wo-photon" coincidence imaging with a classical source.
\newblock {\em Phys. Rev. Lett.}, 89:113601, Aug 2002.

\bibitem{Zhang:05}
D.~Zhang, Y.-H. Zhai, L.-A. Wu, and X.-H. Chen.
\newblock Correlated two-photon imaging with true thermal light.
\newblock {\em Opt. Lett.}, 30(18):2354--2356, Sep 2005.

\bibitem{Genty16}
P.~Ryczkowski, M.~Barbier, A.~T. Friberg, J.~M. Dudley, and G.~Genty.
\newblock Ghost imaging in the time domain.
\newblock {\em Nat. Photon.}, 10:167--170, Mar 2016.

\bibitem{Devaux:16}
F.~Devaux, P.-A. Moreau, S.~Denis, and E.~Lantz.
\newblock Computational temporal ghost imaging.
\newblock {\em Optica}, 3(7):698--701, Jul 2016.

\bibitem{Devaux:17}
F.~Devaux, K.~P. Huy, S.~Denis, E.~Lantz, and P.-A. Moreau.
\newblock Temporal ghost imaging with pseudo-thermal speckle light.
\newblock {\em J. Opt.}, 19(2):024001, 2017.

\bibitem{Lantz2017}
S.~Denis, P.-A. Moreau, F.~Devaux, and E.~Lantz.
\newblock Temporal ghost imaging with twin photons.
\newblock {\em J. Opt.}, 19(3):034002, 2017.

\bibitem{Valencia2003}
G.~Scarcelli, A.~Valencia, S.~Gompers, and Y.~Shih.
\newblock Remote spectral measurement using entangled photons.
\newblock {\em Appl. Phys. Lett.}, 83(26):5560--5562, 2003.

\bibitem{Janassek:18}
Patrick Janassek, S\'ebastien Blumenstein, and Wolfgang Els\"a\ss{}er.
\newblock Ghost spectroscopy with classical thermal light emitted by a
  superluminescent diode.
\newblock {\em Phys. Rev. Applied}, 9:021001, Feb 2018.

\bibitem{Genty:18}
Caroline Amiot, Piotr Ryczkowski, Ari~T. Friberg, J.M. Dudley, and Goery Genty.
\newblock Broadband continuous spectral ghost imaging for high resolution
  spectroscopy.
\newblock {\em arXiv:1805.12424v1}, 2018.

\bibitem{JanassekOL:18}
Patrick Janassek, S\'ebastien Blumenstein, and Wolfgang Els\"a\ss{}er.
\newblock Recovering a hidden polarization by ghost polarimetry.
\newblock {\em Opt. Lett.}, 43:883, 15 Feb 2018.

\bibitem{Shih2005}
M~D'Angelo and YH~Shih.
\newblock Quantum imaging.
\newblock {\em Laser Physics Letters}, 2(12):567--596, 2005.

\bibitem{Shih2007}
Yanhua Shih.
\newblock Quantum imaging.
\newblock {\em IEEE J. Sel. Top. Quant. Electron.}, 13(4):1016--1030, 2007.

\bibitem{Ferri2005}
F.~Ferri, D.~Magatti, A.~Gatti, M.~Bache, E.~Brambilla, and L.~A. Lugiato.
\newblock High-resolution ghost image and ghost diffraction experiments with
  thermal light.
\newblock {\em Phys. Rev. Lett.}, 94:183602, May 2005.

\bibitem{JMOGatti2006}
A.~Gatti, M.~Bache, D.~Magatti, E.~Brambilla, F.~Ferri, and L.~A. Lugiato.
\newblock Coherent imaging with pseudo-thermal incoherent light.
\newblock {\em J. Mod. Opt.}, 53(5-6):739--760, 2006.

\bibitem{MartienssenAJP:64}
Werner Martienssen and Eberhard Spiller.
\newblock Coherence and fluctuations in light beams.
\newblock {\em Am. J. Phys.}, 32:919, 1964.

\bibitem{Arecchi:65}
F.T. Arecchi.
\newblock Measurement of the statistical distribution of gaussian and laser
  sources.
\newblock {\em Phys. Rev. Lett.}, 15:912--916, 1965.

\bibitem{Arecchi:66}
F.T. Arecchi, E.~Gatti, and A.~Sona.
\newblock Time distribution of photons from coherent and gaussian sources.
\newblock {\em Phys. Lett.}, 20:27--29, 1966.

\bibitem{HBT:56}
R.~{Hanbury Brown} and R.~Q. Twiss.
\newblock Correlation between photons in two coherent beams of light.
\newblock {\em Nature}, 177:27, 1956.

\bibitem{Meyers2011}
R.~E. Meyers, K.~S. Deacon, and Y.~Shih.
\newblock Turbulence-free ghost imaging.
\newblock {\em Appl. Phys. Lett.}, 98(11):111115, 2011.

\bibitem{Aspden:15}
R.~S. Aspden, N.~R. Gemmell, P.~A. Morris, D.~S. Tasca, L.~Mertens, M.~G.
  Tanner, R.~A. Kirkwood, A.~Ruggeri, A.~Tosi, R.~W. Boyd, G.~S. Buller, R.~H.
  Hadfield, and M.~J. Padgett.
\newblock Photon-sparse microscopy: visible light imaging using infrared
  illumination.
\newblock {\em Optica}, 2(12):1049--1052, Dec 2015.

\bibitem{LugiatoItalia:13}
L.~A. Lugiato.
\newblock Ghost imaging: Fundamental and applicative aspects.
\newblock {\em Istituto Lombardo (Rend. Scienze)}, 147:139--148, 2013.

\bibitem{Genovese:16}
Marco Genovese.
\newblock Real applications of quantum imaging.
\newblock {\em J. Opt.}, 18:073002, 2016.

\bibitem{Chen_OL:09}
Xi-Hao Chen, Qian Liu, Kai-Hong Luo, , and Ling-An Wu.
\newblock Lensless ghost imaging with true thermal light.
\newblock {\em Opt. Lett.}, 34:695, 2009.

\bibitem{Liu_OL:14}
X.~F. Liu, X.~H. Chen, X.~R. Yao, W.~K. Yu, G.~J. Zhai, and L.~A. Wu.
\newblock Lensless ghost imaging with sunlight.
\newblock {\em Opt. Lett.}, 39:2314, 2014.

\bibitem{Hartmann:17}
S.~Hartmann and W.~Els\"{a}{\ss}er.
\newblock A novel semiconductor-based, fully incoherent amplified spontaneous
  emission light source for ghost imaging.
\newblock {\em Sci. Reports}, 7:41866, 2017.

\bibitem{Boitier:09}
F.~Boitier, A.~Godard, E.~Rosencher, and C.~Fabre.
\newblock Measuring photon bunching at ultrashort timescale by
  two-photon-absorption in semiconductors.
\newblock {\em Nat. Phys.}, 5:267--270, 2009.

\bibitem{Hartmann:15}
S.~Hartmann, A.~Molitor, and W.~Els\"{a}{\ss}er.
\newblock Ultrabroadband ghost imaging exploiting optoelectronic amplified
  spontaneous emission and two-photon detection.
\newblock {\em Opt. Lett.}, 40(24):5770--5773, Dec 2015.

\bibitem{Desurvire:89}
E.~Desurvire and J.R. Simpson.
\newblock Amplification of spontaneous emission in erbium-doped single-mode
  fibers.
\newblock {\em Journal of Lightwave Technology}, 7:835 -- 845, 1989.

\bibitem{Desurvire:94}
E.~Desurvire.
\newblock {\em Erbium-Doped Fiber Amplifiers: Principles and Applications}.
\newblock John Wiley and Sons, New York, 1994.

\bibitem{Arita:08}
Yoshihiko Arita and Paul Ewart.
\newblock {Infra-red multi-mode absorption spectroscopy of acetylene using an
  Er/Yb:glass micro-laser}.
\newblock {\em Optics Express}, 16:4437--4442, 2008.

\bibitem{Wagner:09}
Steven Wagner, Brian T.Fisher, James W.Fleming, and Volker Ebert.
\newblock {TDLAS-based in situ measurement of absolute acetylene concentrations
  in laminar 2D diffusion flames}.
\newblock {\em Proceedings of the Combustion Institute}, 32:839--846, 2009.

\bibitem{Kalashnikov2016}
Dmitry~A. Kalashnikov, Anna~V. Paterova, Sergei~P. Kulik, and et~al.
\newblock Infrared spectroscopy with visible light.
\newblock {\em Nature Photonics}, 10(2):98--101, 2016.

\bibitem{Boitier2013}
F.~Boitier, A.~Godard, N.~Dubreuil, P.~Delaye, C.~Fabre, and E.~Rosencher.
\newblock Two-photon-counting interferometry.
\newblock {\em Phys. Rev. A}, 87:013844, 2013.

\bibitem{Barrow:62}
Gordon~M. Barrow.
\newblock {\em Introduction to molecular spectroscopy}.
\newblock McGraw-Hill, 1962.

\bibitem{Gordon2017}
I.E. Gordon, L.S. Rothman, C.~Hill, R.V. Kochanov, Y.~Tan, P.F. Bernath,
  M.~Birk, V.~Boudon, A.~Campargue, K.V. Chance, B.J. Drouin, J.-M. Flaud, R.R.
  Gamache, J.T. Hodges, D.~Jacquemart, V.I. Perevalov, A.~Perrin, K.P. Shine,
  M.-A.H. Smith, J.~Tennyson, G.C. Toon, H.~Tran, V.G. Tyuterev, A.~Barbe, A.G.
  Cs{\'{a}}sz{\'{a}}r, V.M. Devi, T.~Furtenbacher, J.J. Harrison, J.-M.
  Hartmann, A.~Jolly, T.J. Johnson, T.~Karman, I.~Kleiner, A.A. Kyuberis,
  J.~Loos, O.M. Lyulin, S.T. Massie, S.N. Mikhailenko, N.~Moazzen-Ahmadi,
  H.S.P. M{\"{u}}ller, O.V. Naumenko, A.V. Nikitin, O.L. Polyansky, M.~Rey,
  M.~Rotger, S.W. Sharpe, K.~Sung, E.~Starikova, S.A. Tashkun, J.~Vander
  Auwera, G.~Wagner, J.~Wilzewski, P.~Wcis{\l}o, S.~Yu, and E.J. Zak.
\newblock {The HITRAN2016 molecular spectroscopic database}.
\newblock {\em Journal of Quantitative Spectroscopy and Radiative Transfer},
  203:3--69, dec 2017.

\bibitem{Kuhn:16}
S.~Kuhn, S.~Hartmann, and W.~Els\"{a}{\ss}er.
\newblock Photon-statistics-based classical ghost imaging with one single
  detector.
\newblock {\em Opt. Lett.}, 41(12):2863--2866, Jun 2016.

\bibitem{Hartmann:16}
S.~Hartmann, S.~Kuhn, and W.~Els\"{a}{\ss}er.
\newblock Characteristic properties of the spatial correlations and visibility
  in mixed light ghost imaging.
\newblock {\em Appl. Opt.}, 55(28):7972--7979, Oct 2016.

\bibitem{hauke2011}
Jan Hauke and Tomasz Kossowski.
\newblock Comparison of values of pearson's and spearman's correlation
  coefficients on the same sets of data.
\newblock {\em Quaestiones geographicae}, 30(2):87--93, 2011.

\end{thebibliography}

\end{document}